\input harvmac
\input amssym

\input epsf

\def\unit{\relax{\rm 1\kern-.26em I}}
\def\nada{\relax{\rm 0\kern-.30em l}}



\def\det{{\rm det}}

\noblackbox
\def\IL{\relax{\rm I\kern-.18em L}}
\def\IH{\relax{\rm I\kern-.18em H}}
\def\IR{\relax{\rm I\kern-.18em R}}
\def\IC{\relax\hbox{$\inbar\kern-.3em{\rm C}$}}
\def\IZ{\relax\ifmmode\mathchoice
{\hbox{\cmss Z\kern-.4em Z}}{\hbox{\cmss Z\kern-.4em Z}}
{\lower.9pt\hbox{\cmsss Z\kern-.4em Z}} {\lower1.2pt\hbox{\cmsss
Z\kern-.4em Z}}\else{\cmss Z\kern-.4em Z}\fi}
\def\CM {{\cal M}}
\def\CN {{\cal N}}

\def\CO {{\cal O}}

\def\CC {{\cal C}}

\def\CM {{\cal M}}
\def\CN {{\cal N}}

\def\CO {{\cal O}}

\def\det{{\rm det}}
\def\Tr{{\rm Tr}}

\font\manual=manfnt \def\dbend{\lower3.5pt\hbox{\manual\char127}}

\def\IZ{\relax\ifmmode\mathchoice
{\hbox{\cmss Z\kern-.4em Z}}{\hbox{\cmss Z\kern-.4em Z}}
{\lower.9pt\hbox{\cmsss Z\kern-.4em Z}} {\lower1.2pt\hbox{\cmsss
Z\kern-.4em Z}}\else{\cmss Z\kern-.4em Z}\fi}

\def\lfm#1{\medskip\noindent\item{#1}}

\def\bar{\overline}

\def\rt2{\sqrt{2}}
\def\irt2{{1\over\sqrt{2}}}

\def\tilde{\widetilde}

\def\slashchar#1{\setbox0=\hbox{$#1$}           
   \dimen0=\wd0                                 
   \setbox1=\hbox{/} \dimen1=\wd1               
   \ifdim\dimen0>\dimen1                        
      \rlap{\hbox to \dimen0{\hfil/\hfil}}      
      #1                                        
   \else                                        
      \rlap{\hbox to \dimen1{\hfil$#1$\hfil}}   
      /                                         
   \fi}

\def\foursqr#1#2{{\vcenter{\vbox{
    \hrule height.#2pt
    \hbox{\vrule width.#2pt height#1pt \kern#1pt
    \vrule width.#2pt}
    \hrule height.#2pt
    \hrule height.#2pt
    \hbox{\vrule width.#2pt height#1pt \kern#1pt
    \vrule width.#2pt}
    \hrule height.#2pt
        \hrule height.#2pt
    \hbox{\vrule width.#2pt height#1pt \kern#1pt
    \vrule width.#2pt}
    \hrule height.#2pt
        \hrule height.#2pt
    \hbox{\vrule width.#2pt height#1pt \kern#1pt
    \vrule width.#2pt}
    \hrule height.#2pt}}}}
\def\psqr#1#2{{\vcenter{\vbox{\hrule height.#2pt
    \hbox{\vrule width.#2pt height#1pt \kern#1pt
    \vrule width.#2pt}
    \hrule height.#2pt \hrule height.#2pt
    \hbox{\vrule width.#2pt height#1pt \kern#1pt
    \vrule width.#2pt}
    \hrule height.#2pt}}}}
\def\sqr#1#2{{\vcenter{\vbox{\hrule height.#2pt
    \hbox{\vrule width.#2pt height#1pt \kern#1pt
    \vrule width.#2pt}
    \hrule height.#2pt}}}}


\def\sqd{^2}
\def\nc{N_c}
\def\nf{N_f}

\def\apr{\alpha^\prime}


\lref\KlemmWP{
  A.~Klemm, W.~Lerche and S.~Theisen,
  ``Nonperturbative effective actions of N=2 supersymmetric gauge theories,''
  Int.\ J.\ Mod.\ Phys.\ A {\bf 11}, 1929 (1996)
  [arXiv:hep-th/9505150].
}
\lref\WittenEP{
  E.~Witten,
  ``Branes and the dynamics of {QCD},''
  Nucl.\ Phys.\ B {\bf 507}, 658 (1997)
  [arXiv:hep-th/9706109].
}

\lref\BrandhuberIY{
  A.~Brandhuber, N.~Itzhaki, V.~Kaplunovsky, J.~Sonnenschein and
S.~Yankielowicz,
  ``Comments on the M theory approach to N = 1 S{QCD} and brane dynamics,''
  Phys.\ Lett.\ B {\bf 410}, 27 (1997)
  [arXiv:hep-th/9706127].
}

\lref\SeibergBZ{
  N.~Seiberg,
   ``Exact Results On The Space Of Vacua Of Four-Dimensional Susy Gauge
  Phys.\ Rev.\ D {\bf 49}, 6857 (1994)
  [arXiv:hep-th/9402044].
}
\lref\HitchinZR{
  N.~J.~Hitchin,
  ``Polygons and gravitons,''
In {\it Gibbons, G.W. (ed.), Hawking, S.W. (ed.): Euclidean
quantum gravity} 527-538.
}

\lref\GibbonsNT{
  G.~W.~Gibbons and P.~Rychenkova,
 ``HyperKaehler quotient construction of BPS monopole moduli spaces,''
  Commun.\ Math.\ Phys.\  {\bf 186}, 585 (1997)
  [arXiv:hep-th/9608085].
}

\lref\HananyIE{
  A.~Hanany and E.~Witten,
   ``Type IIB superstrings, BPS monopoles, and three-dimensional gauge
  dynamics,''
  Nucl.\ Phys.\ B {\bf 492}, 152 (1997)
  [arXiv:hep-th/9611230].
}

\lref\FrancoHT{
  S.~Franco, I.~Garcia-Etxebarria and A.~M.~Uranga,
  ``Non-supersymmetric meta-stable vacua from brane configurations,''
  arXiv:hep-th/0607218.
}

\lref\SeibergPQ{
  N.~Seiberg,
  ``Electric - magnetic duality in supersymmetric nonAbelian gauge theories,''
  Nucl.\ Phys.\ B {\bf 435}, 129 (1995)
  [arXiv:hep-th/9411149].
}
\lref\ArgyresEH{
  P.~C.~Argyres, M.~R.~Plesser and N.~Seiberg,
  ``The Moduli Space of N=2 SUSY {QCD} and Duality in N=1 SUSY {QCD},''
  Nucl.\ Phys.\ B {\bf 471}, 159 (1996)
  [arXiv:hep-th/9603042].
}

\lref\HoriAB{
  K.~Hori, H.~Ooguri and Y.~Oz,
  ``Strong coupling dynamics of four-dimensional
N = 1 gauge theories from  M
  theory fivebrane,''
  Adv.\ Theor.\ Math.\ Phys.\  {\bf 1}, 1 (1998)
  [arXiv:hep-th/9706082].
}

\lref\WittenNii{
 E.~Witten, ``Solutions of Four-Dimensional Field Theories Via
 M-Theory," [arXiv:hep-th/9703166]
 }

\lref\deBoerZY{
  J.~de Boer, K.~Hori, H.~Ooguri and Y.~Oz,
  ``Kaehler potential and higher derivative terms from M theory five-brane,''
  Nucl.\ Phys.\ B {\bf 518}, 173 (1998)
  [arXiv:hep-th/9711143].
}

\lref\ElitzurFH{
  S.~Elitzur, A.~Giveon and D.~Kutasov,
  ``Branes and N = 1 duality in string theory,''
  Phys.\ Lett.\ B {\bf 400}, 269 (1997)
  [arXiv:hep-th/9702014].
}

\lref\SchmaltzSQ{
  M.~Schmaltz and R.~Sundrum,
  ``N = 1 field theory duality from M-theory,''
  Phys.\ Rev.\ D {\bf 57}, 6455 (1998)
  [arXiv:hep-th/9708015].
}

\lref\HoriIW{
  K.~Hori,
  ``Branes and electric-magnetic duality in supersymmetric {QCD},''
  Nucl.\ Phys.\ B {\bf 540}, 187 (1999)
  [arXiv:hep-th/9805142].
}

\lref\GiveonSR{
  A.~Giveon and D.~Kutasov,
   ``Brane dynamics and gauge theory,''
  Rev.\ Mod.\ Phys.\  {\bf 71}, 983 (1999)
  [arXiv:hep-th/9802067].
}
\lref\iss{  K.~Intriligator, N.~Seiberg and D.~Shih,
  ``Dynamical SUSY breaking in meta-stable vacua,''
  JHEP {\bf 0604}, 021 (2006)
  [arXiv:hep-th/0602239].
}

\lref\CallanKZ{
  C.~G.~Callan and J.~M.~Maldacena,
  ``Brane dynamics from the Born-Infeld action,''
  Nucl.\ Phys.\ B {\bf 513}, 198 (1998)
  [arXiv:hep-th/9708147].
}

\lref\nsact{
  I.~A.~Bandos, A.~Nurmagambetov and D.~P.~Sorokin,
  ``The type IIA NS5-brane,''
  Nucl.\ Phys.\ B {\bf 586}, 315 (2000)
  [arXiv:hep-th/0003169].
}

\lref\KlebanovHB{
  I.~R.~Klebanov and M.~J.~Strassler,
   ``Supergravity and a confining gauge theory: Duality cascades and
  $\chi$SB-resolution of naked singularities,''
  JHEP {\bf 0008}, 052 (2000)
  [arXiv:hep-th/0007191].
}

\lref\OoguriBG{
  H.~Ooguri and Y.~Ookouchi, ``Meta-Stable Supersymmetry Breaking
  Vacua on Intersecting Branes,'' arXiv:hep-th/0607183.
}

\lref\IntriligatorAU{
  K.~A.~Intriligator and N.~Seiberg,
  ``Lectures on supersymmetric gauge theories and electric-magnetic  duality,''
  Nucl.\ Phys.\ Proc.\ Suppl.\  {\bf 45BC}, 1 (1996)
  [arXiv:hep-th/9509066].
}

\lref\SeibergPQ{
  N.~Seiberg,
  ``Electric - magnetic duality in supersymmetric nonAbelian gauge theories,''
  Nucl.\ Phys.\ B {\bf 435}, 129 (1995)
  [arXiv:hep-th/9411149].
}

\lref\BerkoozKM{
  M.~Berkooz, M.~R.~Douglas and R.~G.~Leigh,
  ``Branes intersecting at angles,''
  Nucl.\ Phys.\ B {\bf 480}, 265 (1996)
  [arXiv:hep-th/9606139].
}

\lref\tdual{
  M.~Bershadsky, C.~Vafa and V.~Sadov,
  ``D-Strings on D-Manifolds,''
  Nucl.\ Phys.\ B {\bf 463}, 398 (1996)
  [arXiv:hep-th/9510225].
  A.~M.~Uranga,
  ``Brane configurations for branes at conifolds,''
  JHEP {\bf 9901}, 022 (1999)
  [arXiv:hep-th/9811004].
  K.~Dasgupta and S.~Mukhi,
  ``Brane constructions, conifolds and M-theory,''
  Nucl.\ Phys.\ B {\bf 551}, 204 (1999)
  [arXiv:hep-th/9811139].
  A.~Giveon, D.~Kutasov and O.~Pelc,
  ``Holography for non-critical superstrings,''
  JHEP {\bf 9910}, 035 (1999)
  [arXiv:hep-th/9907178].
}


\lref\FrancoES{
  S.~Franco and A.~M.~Uranga,
   ``Dynamical SUSY breaking at meta-stable minima from D-branes at obstructed
  geometries,''
  JHEP {\bf 0606}, 031 (2006)
  [arXiv:hep-th/0604136].
}

\lref\OoguriPJ{
  H.~Ooguri and Y.~Ookouchi,
   ``Landscape of supersymmetry breaking vacua in geometrically realized gauge
 theories,''
  arXiv:hep-th/0606061.
}

\lref\ForsteZC{
  S.~Forste,
  ``Gauging flavour in meta-stable SUSY breaking models,''
  arXiv:hep-th/0608036.
}
\lref\BraunDA{
  V.~Braun, E.~I.~Buchbinder and B.~A.~Ovrut,
  ``Towards realizing dynamical SUSY breaking in heterotic model building,''
  arXiv:hep-th/0606241.
}

\lref\AmaritiVK{
  A.~Amariti, L.~Girardello and A.~Mariotti,
  ``Non-supersymmetric meta-stable vacua in SU(N) SQCD with adjoint matter,''
  arXiv:hep-th/0608063.
}

\lref\RayWK{
  S.~Ray,
   ``Some properties of meta-stable supersymmetry-breaking vacua in Wess-Zumino
  models,''
  arXiv:hep-th/0607172.
}

\lref\BanksMA{
  T.~Banks,
  ``Remodeling the pentagon after the events of 2/23/06,''
  arXiv:hep-ph/0606313.
}

\lref\BraunEM{
  V.~Braun, E.~I.~Buchbinder and B.~A.~Ovrut,
  ``Dynamical SUSY breaking in heterotic M-theory,''
  Phys.\ Lett.\ B {\bf 639}, 566 (2006)
  [arXiv:hep-th/0606166].
}


\Title{\rightline{hep-th/0608157} } {\vbox{\centerline{A Note on
(Meta)stable}
\smallskip \centerline{ Brane Configurations in MQCD} }}
\medskip
\centerline{I. Bena$^{(1)}$, E. Gorbatov$^{(2)}$, S.
Hellerman$^{(1)}$, N. Seiberg$^{(1)}$, and D. Shih$^{(3)}$}

\bigskip
\centerline{{${}^{(1)}$\it School of Natural Sciences, Institute
for Advanced Study }} \centerline{{\it Einstein Dr., Princeton, NJ
08540}}
\medskip
\centerline{{${}^{(2)}$\it Department of Physics, University of
California San Diego}} \centerline{{\it 9500 Gilman Drive, La
Jolla, CA 92093}}
\medskip
\centerline{{${}^{(3)}$\it Department of Physics, Princeton
University }} \centerline{{\it Princeton, NJ 08540}}
\medskip
\bigskip

\noindent We examine the M-theory version of SQCD which is known
as MQCD. In the IIA limit, this theory appears to have a
supersymmetry-breaking brane configuration which corresponds to
the meta-stable state of $\CN=1$ $SU(N_c)$ SQCD. However, the
behavior at infinity of this non-supersymmetric brane construction
differs from that of the supersymmetric ground state of MQCD.  We
interpret this to mean that it is not a meta-stable state in MQCD,
but rather a state in another theory.  This provides a concrete
example of the fact that, while MQCD accurately describes the
supersymmetric features of SCQD, it fails to reproduce its
non-supersymmetric features (such as meta-stable states) not only
quantitatively but also qualitatively.

\vskip 0.4in \Date{August  2006}

\newsec{Introduction}

Stringy versions of quantum field theories have proved to be an
important tool in giving an intuitive geometric understanding of
gauge theory dynamics. In particular, one can construct
supersymmetric configurations of D-branes and NS5 branes in type
IIA string theory whose low energy spectrum matches that of a wide
variety of four-dimensional supersymmetric gauge theories
\refs{\HananyIE\WittenNii-\GiveonSR}. We will refer henceforth to
the general, nonrenormalizable type IIA stringy embeddings of
these field theories
as the
parameter space of MQCD theories. For generic values of the
parameters, the correct description of MQCD is in terms of objects
in M-theory, such as M5 branes. But MQCD also has two interesting
$g_s\to 0$ limits that we will consider in this paper.

The first limit is when $g_s$ is infinitesimal and all the length scales
of the system are larger than $\sim\ell_s$. This we will call the
``D-brane limit," because in this limit the dynamics of the system
may usefully be described by the Born-Infeld action of a set of
D-branes moving in a flat background.  The interaction of the
dynamical D-branes of interest with other background branes is
given by boundary conditions on the dynamical D-branes, and by the
introduction of additional, bifundamental degrees of freedom
charged under various gauge fields. The degrees
of freedom in this limit are the open strings where both $g_s$ and
$\apr$ corrections can be treated as small.

The second limit is one in which $g_s$ is infinitesimal, but
various length scales also scale to zero as $g_s\ell_s$. This we
will refer to as the ``decoupling limit" or ``SQCD limit,'' because
in this limit one obtains an interacting gauge theory on the
D-branes, decoupled from all other stringy degrees of freedom.

For theories with ${\cal N} = 1$ supersymmetry in 4 dimensions, the
decoupling limit and D-brane limit of MQCD theories generally do
not overlap. In general the string embeddings are coupled to a
large number of modes above and beyond the modes of the 4D field
theory one wants to study. Bulk gravity modes, Kaluza-Klein modes
of higher-dimensional field theories, light modes on nearby
branes, string oscillator modes and other degrees of freedom
interact with the degrees of freedom of the 4D field theory of
interest. In the D-brane limit, these massive modes have not
decoupled, so the field theory description is not valid.
Conversely, when the field theory description is applicable, one
cannot view the system in terms of classical D-branes, since these
branes are extended over sub-stringy distances.

Although these two limits lead to different theories, holomorphy
ensures that supersymmetric quantities calculated both in the two limits or in
other points of the MQCD parameter space
can be smoothly continued to any other point of the  MQCD
parameter space \refs{\WittenEP\HoriAB-\BrandhuberIY}. On the
other hand, non-supersymmetric quantities are not protected by
holomorphy, and so they cannot in principle be continued from one limit
to another. Nevertheless,  one might have hoped that
non-supersymmetric features of the dynamics should, at least
qualitatively, be independent of one's precise location in the
parameter space of MQCD. In this paper, we will find a concrete
counterexample to this, in the specific context of $\CN=1$
$SU(N_c)$ SQCD and its extension to MQCD.

Recently, it has been shown that massive $\CN=1$ $SU(N_c)$ SQCD in the
free-magnetic phase exhibits dynamical supersymmetry breaking in
meta-stable vacua \iss. This phenomenon was subsequently demonstrated
in many similar theories
\refs{\FrancoES\BraunDA\OoguriPJ\BraunEM\BanksMA\RayWK\ForsteZC-\AmaritiVK}.  One might hope to find such
meta-stable vacua at other points in the MQCD parameter space.
Instead, we will show in this paper that there is a simple obstruction
to continuing the meta-stable vacuum of SQCD into MQCD.

The obstruction has to do with the phenomenon of brane bending. At
tree-level in $g_s$, the NS5-branes in the brane construction are
straight. However, because of string interactions they curve
\WittenNii. The direction of this bending is determined by the charges and directions of the D-branes which end on the NS5-brane. Therefore,
a proper way to {\it define}
the theory on the branes at $g_s \neq 0$ is not in terms of the detailed positions of the
branes, but instead, in terms of the asymptotic behavior of the
branes that stretch to infinity. These are the boundary
conditions on the system, while the branes in the interior are
dynamical and free to adjust themselves. The supersymmetric brane
configuration leads to a specific bending of the NS5-branes, and
this bending defines a point in the MQCD parameter space. Every
state of this theory, stable or meta-stable must have these
boundary conditions at infinity.  If it has different boundary
conditions it is not a state of the same system.

Now that we understand how to define the system using its correct
asymptotic behavior, we can check how the nonsupersymmetric brane
configuration bends at infinity.  The main result of our
investigation is that, at $g_s \neq 0$, the putative
nonsupersymmetric brane configuration cannot bend at infinity
according to the proper boundary conditions. The only solutions of
the minimal area equations (or of the equations coming from the
NS5 brane worldvolume action) bend in the wrong direction. Hence,
they differ by an infinite amount from the supersymmetric brane
configurations. These solutions therefore do not correspond to
meta-stable MQCD vacua.

Our results indicate that qualitative nonsupersymmetric features
of SQCD, like the existence of a meta-stable vacuum, are not seen
in the brane description of this theory. While our analysis does
not prove that such meta-stable vacua cannot be seen in brane
constructions of other four-dimensional gauge theories, the fact
that the simplest example of a gauge theory meta-stable vacuum
does not carry over to MQCD seems to indicate that this problem is
generic.

It would be most interesting to explore if these kinds of
meta-stable vacua exist in other string-theoretic dual
descriptions of gauge theories, like the gauge-gravity duality.
Another interesting application is to ``T-dual'' versions of the
brane constructions. These dualize to branes near a Calabi-Yau
singularity that preserves $\CN=1$ supersymmetry \tdual.
It would be interesting to see
whether the subtleties we uncover here also appear in this description.

The outline of our paper is as follows. We begin in section 2 by
reviewing the old brane constructions of massless $\CN=1$ $SU(N_c)$
SQCD and its magnetic dual. The two brane constructions are obtained
as two singular limits of the parameter space of MQCD (see figure
1). Different nonoverlapping regions of the diagram are associated
with different simple descriptions of the system. The decoupling limit
corresponds to the immediate vicinity of the origin where $\Delta L\to
0$. The magnetic brane configuration is obtained for fixed and
positive value of $\Delta L$, while the electric brane configuration
is obtained for fixed and negative $\Delta L$.

\midinsert \vskip .2cm \centerline{ {\epsfxsize
4.0in\epsfbox{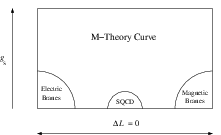}}} \vskip .25cm \leftskip 2pc \rightskip
2pc\noindent{\ninepoint\sl \baselineskip=8pt {\bf Figure 1}:  A
schematic description of the MQCD parameter space for $m = 0$.  At
small $g_ s$, there are two ``brane limits'' -- one where $\Delta
L$ is fixed and negative, another where $\Delta L$ is fixed and
positive as $g_s \to 0$. The SQCD decoupling limit is obtained by
taking $\Delta L \to 0$ and $g_ s \to 0$ with $g_{YM}^2\sim
g_s\ell_s/\Delta L$ fixed. These three limits do not overlap with
one another. Since the 4D gauge coupling vanishes in both brane
limits, there can be no duality connecting them directly. }
\endinsert

We then go on to consider the system at nonzero mass. Here the
situation is quite different and is shown in figure 2 -- turning
on a mass yields a SUSY brane configuration in the electric phase
but not in the magnetic phase. This parallels the fact that there
is a tree-level SUSY vacuum in massive SQCD, but not in its
magnetic dual. In the magnetic phase (as in the field theory), the
system becomes tachyonic. We follow the tachyon condensation and
obtain a pseudo-moduli space of minimal-energy, SUSY-breaking
brane configurations. We show that these brane configurations have
the right properties to correspond to the tree-level SUSY-breaking
vacua of the magnetic dual to massive SQCD \SeibergPQ.

\midinsert \vskip .2cm \centerline{ {\epsfxsize
4.0in\epsfbox{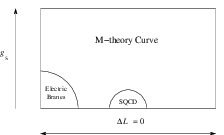}}} \vskip .25cm \leftskip 2pc \rightskip
2pc\noindent{\ninepoint\sl \baselineskip=8pt {\bf Figure 2}:  A
schematic description of the MQCD parameter space for $m \ne 0$.
At small $g_s$, there is only one ``brane limit'' -- the one where
$\Delta L$ is fixed and negative.
 The SQCD decoupling limit is again
obtained by taking $\Delta L \to 0$ and $g_s \to 0$ with $g_
{YM}\sqd$ fixed. }
\endinsert

In section 3 we analyze the possibility of lifting these
non-supersymmetric configurations to M-theory. To set up the problem,
we first review known results on M-theory lifts of the supersymmetric
brane configurations. These lift to M5 branes wrapping holomorphic
curves in Taub-NUT$\times {\Bbb R}^2$. We describe how at zero mass,
both the electric and magnetic brane setups lift to the same
holomorphic curve in M-theory. Equivalently, we can take two different
limits of the same curve in M-theory and find the two different brane
setups
\refs{\SchmaltzSQ,\HoriIW}. This is the sense in which the
electric-magnetic duality of \SeibergPQ\ is implemented in this
context. At nonzero mass, we explain how a smooth supersymmetric
M-theory curve exists for every nonzero value of the mass parameter,
but -- unlike the situation for zero mass -- there is no smooth limit
which corresponds to a magnetic brane configuration, and only the
limit to the electric brane configuration is well-defined (see figure
2).

We then tackle the problem of lifting the tree-level SUSY-breaking
brane configurations of section 2 to M-theory. Such a
configuration should lift to M5 branes wrapping  non-holomorphic
minimal-area curves in Taub-NUT$\times {\Bbb R}^2$. We solve
the (rather involved) equations of motion for these curves, and
show that there is no solution
with the same boundary conditions as the SUSY vacua. In section 4 we
undertake a perturbative type IIA analysis of the SUSY-breaking
brane configuration at small $g_s$, and we find the same result.
We conclude from this that, after string interactions are taken
into account, there is no meta-stable brane configuration in
the D-brane limit of MQCD.

In order to further understand the absence of the metastable vacuum,
we analyze in section 5 a kink
``quasi-solution'' that has the right asymptotic boundary
conditions, and reduces to the SUSY-breaking configuration at $g_s
\to 0$. We find that this brane ``quasi-solution'' has a runaway
mode, which ruins the stability of the solution at nonzero $g_s$.
Only in the $g_s \to 0$ limit is this runaway mode frozen, and
this is the reason the $g_s = 0$ SUSY-breaking brane
configuration is stable. Finally, in appendix A we include some
technical details regarding the exact solution of the M5 brane
equations of motion.

\medskip

\noindent{\bf Note:} While this paper was being prepared for
publication, two papers which explore the SUSY-breaking brane
configuration appeared \refs{\OoguriBG,\FrancoHT}.  These
results overlap with parts of section 2 of our paper. The authors
of \FrancoHT\ also propose a lift of this brane configuration to
M-theory, which, as we discuss in more detail in section 3, has
different boundary conditions and hence is not a meta-stable MQCD
state.

\newsec{ IIA Brane Configurations with $g_s\to 0$}

In this section we review the IIA brane construction of $SU(N_c)$
SQCD and its magnetic dual. Along the way we will clarify a few
points that we found confusing in the literature. More details can
be found in the review \GiveonSR\ and references therein.

The brane constructions described in this section ignore the
phenomenon of brane bending described in the introduction. As
such, they are only capable of describing tree-level physics.
To see any effects of string interactions,
one must go to M-theory, which is the subject of the next section.

\subsec{Electric configuration}

One can realize massless $\CN=1$ SQCD with $N_f$ flavors and gauge
group $SU(N_c)$ on the following set of branes \ElitzurFH:

\lfm{$\bullet$} $ N_f$ D6 branes stretched in the 0123789
directions, at $x^{4,5,6}=0$.

\lfm{$\bullet$} An NS5 brane in the 012345 directions at
$x^{8,9}=0$,
  $x^6=L_0+\Delta L$, with $\Delta L<0$ (the NS brane).

\lfm{$\bullet$} An NS5 brane in the 012389 directions at
$x^{4}=x^5=0$,
 $x^6=L_0$ (the NS' brane).

\lfm{$\bullet$} $N_f$ D4 branes running between the D6 brane and
the NS brane (the ``flavor" D4 branes).

\lfm{$\bullet$} $N_c $ D4 branes running between the NS' brane and
 the NS brane (the ``color" D4 branes).

\medskip

\noindent This brane configuration is depicted in figure 3.

Let us briefly describe the low-energy effective field theory
living on the D4 branes. The massless modes comprise $\CN=1$
$SU(N_c)$ SQCD: the 4-4 strings between the color branes give rise
to 4D $SU(N_c)$ gauge fields; while the 4-4 strings between the
color and flavor branes give rise to $N_f$ quarks, massless fields
$Q_f$, $\tilde Q_f$ transforming in the fundamental of $SU(N_c)$.
Note that the quarks are four-dimensional because the 4-4 strings
are localized at the intersection of the color and flavor branes.

Of course, the theory on the branes is not precisely $\CN=1$ SQCD
-- there are infinite towers of massive modes with masses $\sim
|\Delta L|^{-1}$, $L_0^{-1}$ coming from Kaluza-Klein reduction of
the various 4-4 strings, as well as the usual open and closed
string states with masses $\sim \ell_s^{-1}$. In particular,
notice that the 4-4 open strings connecting the flavor branes are
all massive (with masses $\sim L_0^{-1}$) because of the boundary
conditions at the D6 and the NS brane.

\centerline{\epsfxsize=0.50\hsize\epsfbox{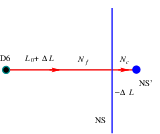}}
\noindent{\ninepoint\sl \baselineskip=8pt {\bf Figure 3}:{\sl $\;$
The electric brane configuration describing $SU(N_c)$ $\CN=1$
SQCD. The axes are as follows: the vertical direction represents
$(x^4,\,x^5)$, the horizontal direction is $x^6$, and coming out
of the page are $(x^8,\,x^9)$. All other directions (including the
non-compact field theory directions) are suppressed. }}
\medskip

If the goal is to decouple the massive states and obtain precisely
$SU(N_c)$ SQCD, we should send $\Delta L$, $L_0$ and $\ell_s\to 0$
holding some field theory energy (such as the dynamical scale
$\Lambda$) fixed. More precisely, in order to obtain a theory with
finite 4D gauge coupling, we should take
\eqn\gszlim{
g_s\to 0,\qquad \ell_s \to 0,\qquad \Delta L\to 0
}
with
\eqn\lambdadef{
{1\over g_{elec}^2} ={|\Delta L|\over g_s\ell_s}
}
held fixed. This leaves behind $\CN=1$ $SU(N_c)$ SQCD, with gauge
coupling $g=g_{elec}$.

However, these scaling limits should be interpreted with care. In
the limit \gszlim, the color D4 branes become extended by only a
sub-stringy amount $\Delta L\sim g_s\ell_s$ deep within the throat
of the NS5 branes. So at some point in the decoupling limit, the
classical brane picture will break down, and along with it the
naive relation \lambdadef.\foot{In particular, we expect that
\lambdadef\ will receive logarithmic renormalization $\sim
\log(|\Delta L|\Lambda)$, which becomes important as $\Delta L\to
0$.} But at that point, we should switch over to the description
in terms of the decoupled field theory. The upshot is that there
is never a regime of parameters where both the flat space string
theory and the field theory are simultaneously valid. As discussed
in the introduction, we will focus in this paper on the regime of
parameters where the string theory description is valid. In
particular, all the length scales ($L_0$, $\Delta L$, $\ell_s$
etc.) will be order one, and massive modes will not have
decoupled.

Finally, let us conclude our review of the electric brane
configuration by describing a deformation that will be relevant
below: moving the D6 branes by an amount $\Delta x$ in the
$x^4+ix^5$ direction. Shown in figure 4 is the supersymmetric
configuration resulting from this deformation. As far as the
massless modes on the D4 branes are concerned, this deformation
corresponds to deforming the superpotential by a mass term for the
quarks:\foot{Notice also that there is only one SUSY vacuum here;
the $N_c$ SUSY vacua one expects from the Witten index are only
visible after nonperturbative quantum effects are taken into
account. This can be done either by studying the gauge theory on the
D4 branes, or by going to the M-theory lift.} \eqn\Welec{ W = \Tr\,m
Q\tilde Q } The relation between $m$ and $\Delta x$ in the classical
brane picture (where one neglects the background geometry sourced by
the NS5 branes) is \eqn\melec{ m = {\Delta x\over\ell_s^2} } Again,
we expect this naive relation to be modified in the field theory
decoupling limit \gszlim.

\centerline{\epsfxsize=0.50\hsize\epsfbox{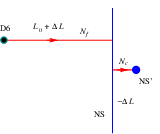}}
\noindent{\ninepoint\sl \baselineskip=8pt {\bf Figure 4}:{\sl $\;$
The brane configuration describing the SUSY vacuum of massive
$\CN=1$ SQCD. The axes are the same as in figure 3.
}}

\subsec{Magnetic configuration -- massless case}

One can realize the magnetic dual to massless $\CN=1$ $SU(N_c)$
SQCD \SeibergPQ\ using the following set of branes \ElitzurFH:

\lfm{ $\bullet$} $ N_f$ D6 branes stretched in the 0123789
directions, at $x^{4,5,6}=0$.

\lfm{ $\bullet$} An NS5 brane in the 012389 directions at
$x^{4}=x^5=0$, $x^6=L_0$ (the NS' brane).

\lfm{  $\bullet$} An NS5 brane in the 012345 directions at
$x^{8,9}=0$, $x^6=L_0+\Delta L$, with $\Delta L > 0$ (the NS
brane)

\lfm{  $\bullet$} $\nf$ D4 branes running between the D6 brane and
the NS' brane (the ``flavor" branes).

\lfm{ $\bullet$} $\nf - \nc $ D4 branes running between the NS'
brane and the NS brane (the ``color" branes).

\medskip

\noindent This brane configuration is shown in figure 5. In
particular we should note that it has {\it different} behavior at
infinity compared to the electric brane configuration shown in figure
3. Therefore, it describes a different theory in the parameter space
of MQCD.

\centerline{\epsfxsize=0.40\hsize\epsfbox{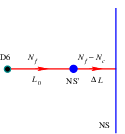}}
\noindent{\ninepoint\sl \baselineskip=8pt {\bf Figure 5}:{\sl $\;$
The brane configuration describing the magnetic dual to massless
$\CN=1$ $SU(N_c)$ SQCD. The axes are the same as in figure 3. }}
\bigskip

Now the low-energy effective field theory living on the D4-branes
is given by the following. The massless fields are those of the
magnetic dual to SQCD \SeibergPQ: 4-4 strings between the color branes give
rise to 4D $SU(N_f-N_c)$ gauge fields; 4-4 strings between the
color and flavor branes give rise to the magnetic quarks, i.e.\
fields $q_f$, $\tilde q_f$ transforming in the fundamental of
$SU(N_f-N_c)$; and fluctuations of the 4-4 flavor strings in the
$(x^8,x^9)$ directions give rise to an $N_f\times N_f$ matrix of
gauge-singlet chiral superfields
\eqn\wdef{
w_{fg}=(x^8+ix^9)_{fg}
}
Our conventions are that $q$ and $\tilde q$ are canonically
normalized, dimension one fields, while $w$ has mass dimension -1.
These fields have leading-order K\"ahler potential given by
\eqn\Kw{
K = {L_0\over g_s\ell_s^5}w^\dagger w+q^\dagger q+\tilde q^\dagger
\tilde q+\dots
}
coming from the Born-Infeld action on the flavor D4 branes. From
the geometry, it is clear that the massless fields interact via
the superpotential
\eqn\Wmaggeom{
W ={1\over\ell_s^2}\Tr\, wq\tilde q
}
where the Yukawa coupling is fixed by the requirement that the
magnetic quarks have physical mass $m_q= \Delta w/\ell_s^2$ when
the flavor branes are moved by an amount $\Delta w$ in the $x^8+ix^9$
direction.

Again, there are other modes with masses $\sim (\Delta L)^{-1}$,
$L_0^{-1}$ and $\ell_s^{-1}$ which are not decoupled. If we want
to remove these states, we should send
\eqn\gszlimmag{
g_s\to 0,\qquad \ell_s \to 0,\qquad \Delta L\to 0,\qquad L_0\to 0
}
with
\eqn\fixedmag{
{1\over g_{mag}^2}= {\Delta L\over g_s\ell_s},\qquad
h=\sqrt{g_s\ell_s\over L_0}
}
held fixed. This leaves behind the magnetic dual to $\CN=1$
$SU(N_c)$ SQCD  \SeibergPQ, with gauge coupling $g=g_{mag}$ and
Yukawa coupling
$h$. (Note that the Yukawa coupling $h$ is defined in terms of the
canonically normalized field $\Phi=\sqrt{L_0\over g_s\ell_s^5}w$.)

As was the case for the electric brane configuration, the scaling
limits should again be interpreted with care. In the limit
\gszlimmag, we again expect that the classical brane picture will
break down, and the relations \fixedmag\ will be modified.
Nevertheless, it is interesting that in the decoupling limits
\gszlim\ and \gszlimmag, $\Delta L\to 0$, so the electric and
magnetic brane configurations appear to become indistinguishable
at infinity. This is a hint of electric-magnetic duality, but it
is far from a proof, since the brane configurations cannot be
trusted in this limit.

\subsec{Magnetic configuration -- massive case}

In the magnetic brane configuration, we can also consider moving the
D6 branes by an amount $\Delta x$ in the $x^4+ix^5$ direction. This
has a very different effect than in the electric brane configuration
-- it breaks supersymmetry, as there is no arrangement of the D4
branes, satisfying charge conservation on the NS5 and D6 branes,
which is supersymmetric. The brane configuration one obtains from
this deformation is shown in figure 6. There is an open string
tachyon between the color and flavor branes, with mass squared
 \eqn\mtachmag{ m_{tach}^2= -{\Delta x \over \ell_s^2L_0} }
Again, we stress that this formula is only valid in the classical
brane picture, away from the decoupling limit.

\centerline{\epsfxsize=0.40\hsize\epsfbox{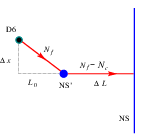}}
\noindent{\ninepoint\sl \baselineskip=8pt {\bf Figure 6}:{\sl $\;$
The magnetic brane configuration describing the origin of moduli
space in $N_f>N_c$ massive $\CN=1$ SQCD. }}
\medskip

In the field theory living on the D4 branes, the deformation by
$\Delta x$ corresponds to adding a term linear in $w$ to the
superpotential:
\eqn\Wmaglin{
W = {1\over\ell_s^2}\Tr\,w q\tilde q + {\Delta x \over
g_s\ell_s^5}\Tr\,w
}
where the coefficient of the linear term is fixed by matching the
mass of the field theory tachyon with that of the open string
tachyon \mtachmag. Indeed, the $w$ F-term contributes a term to
the scalar potential of the form
\eqn\DeltaVST{\eqalign{
   V \supset K_{w\bar w}^{-1}|F_w|^2 &=\left( {L_0\over
g_s\ell_s^5}\right)^{-1}\Tr\left| {1\over\ell_s^2}q\tilde
q+{\Delta x \over g_s\ell_s^5}\right|^2\cr
 & ={N_f|\Delta x|^2\over g_s\ell_s^5 L_0} + {\Delta x\over \ell_s^2 L_0} (q\tilde
q+c.c.) + \CO((q\tilde q)^2)
}}
We see in the $q\tilde q $ quadratic term precisely the open string
tachyon with the requisite mass.

The linear term in \Wmaglin\ spontaneously breaks supersymmetry at
tree level (and to all orders in perturbation theory) by the rank
condition. It is important that the supersymmetry breaking must be
spontaneous in the field theory, because the brane configuration
has supersymmetric boundary conditions -- the only objects that
stretch to infinity are the mutually BPS NS5 branes and D6 brane.
Notice also that the gauge theory on the branes is expected to
have $N_c$ SUSY vacua coming from non-perturbative effects in the
magnetic gauge group. We cannot see these here, just as we could
not see the $N_c$ SUSY vacua in the electric brane configuration.
As discussed at the beginning of the section, the reason is that
the brane configuration only captures the physics that is
perturbative in $g_s$. Non-perturbative physics is only visible in
the M-theory lift, to be discussed in the next section.

As a consistency check, we can compute the vacuum energy density
of this (tachyonic) state two ways, first using field theory
\Wmaglin, which gives the constant piece in \DeltaVST:
\eqn\vactach{
V_{tach} ={N_f|\Delta x|^2\over g_s\ell_s^5 L_0}
}
and second with the DBI action, i.e.\ the length of the D4 branes,
\eqn\vacdbi{
V_{DBI}=\tau_4\left(N_f\sqrt{|\Delta x|^2+L_0^2}+(N_f-N_c)|\Delta
L|\right)
 }
In the $\Delta x\ll L_0$ limit, and substituting $\tau_4\sim
1/g_s\ell_s^5$ for the D4 brane tension, we find agreement between
\vactach\ and the $|\Delta x|$ dependent part of \vacdbi. This is
a consistency check of the approximately canonical K\"ahler potential
\Kw.

\subsec{Minimal-energy SUSY-breaking brane configuration}

Since the configuration shown in figure 6 is tachyonic, it can lower
its energy by condensing the tachyon. This happens by reconnecting
$N_f-N_c$ of the D4 branes, whereby they snap into the horizontal
$x^6$ direction. The resulting minimal-energy SUSY-breaking
configuration is shown in figure 7.

\goodbreak\midinsert
\vskip -.2cm \centerline{ {\epsfxsize
2.3in\epsfbox{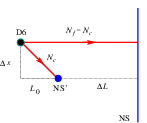}}} \vskip -.25cm \leftskip 2pc
\rightskip 2pc\noindent{\ninepoint\sl \baselineskip=8pt {\bf
Figure 7}: The minimal-energy SUSY-breaking brane configuration. }
\endinsert

In the field theory on the D4 branes, the snapping of the D4 branes
corresponds to giving $q\tilde q$ an expectation value $\sim \Delta
x$, which Higgses the magnetic gauge group. Indeed, since the
superpotential and leading-order K\"ahler potential are
algebraically identical to the ones analyzed in \iss, one can repeat
the calculations in that paper line by line to show that the system
described by \Kw\Wmaglin\ has a SUSY-breaking vacuum, in which $(\nf
-\nc)^2$ of the $q$'s and $\tilde q$'s have acquired an expectation
value.

The brane configuration described in figure 7 is clearly the
analogue of the SUSY-breaking vacuum found at {\it tree-level} in
the magnetic dual to SQCD.\foot{We should stress, however, the
central point of this paper: due to string interactions, this
brane configuration is actually {\it not} related to the
meta-stable state of SQCD. We will elaborate on this in detail in
the later sections.} In the remainder of this subsection, we will
list some consistency checks of this claim.

{\it (1) Energies and energy differences.} One obvious check is to
compare the energy of the SUSY-breaking vacuum, $V_{\rm 0}$,
computed using the effective field theory and using the DBI
action. As in \iss, the former gives
\eqn\Vmin{
V_0 ={N_c|\Delta x|^2\over g_s\ell_s^5 L_0}
}
while the latter gives
\eqn\vacdbiii{
V_{DBI}=\tau_4\left(N_c\sqrt{|\Delta
x|^2+L_0^2}+(N_f-N_c)(L_0+|\Delta L|)\right)
}
Again, we find complete agreement of the $|\Delta x|$ dependent
part, in the limit $|\Delta x|\ll L_0$. We can also compare the
energy difference between the tachyonic state and the vacuum:
\eqn\vacdbiiii{
\Delta V_{DBI} = \tau_4(N_f-N_c)\left( \sqrt{|\Delta
x|^2+L_0^2}-L_0\right)
}
which agrees exactly with the energy difference computed from the
field theory in the $\Delta x\ll L_0$ limit.

{\it (2) Symmetries.} The SUSY-breaking brane configuration
carries the same quantum numbers as the SQCD meta-stable state.
First of all, there are no D4 branes between the NS5 branes, so
the magnetic gauge group is completely Higgsed. Second, there are
two groups of D4 branes, and so the non-Abelian flavor symmetry
$SU(N_f)$ is spontaneously broken down to $SU(N_f-N_c)\times
SU(N_c)$, as in SQCD. Finally, the configuration is invariant
under rotations in the 89 plane, so the state preserves the same
accidental R-symmetry as in SQCD, under which the mesons have
charge 2 and the magnetic quarks charge 0. (For a discussion of
the various $U(1)$ factors in the brane configuration and how they match
onto the field theory, see e.g.\ \GiveonSR, p. 153.)

{\it (3) Pseudo-moduli.} The SUSY-breaking brane configuration
also has the same non-compact pseudo-moduli as the SQCD
meta-stable state. Since the D6 and the NS' are parallel, we can
slide the $N_c$ D4 branes independently along the 89 directions.
These modes correspond to the eigenvalues of the $N_c\times N_c$
matrix of pseudo-moduli $\delta\Phi_0$ in the field theory.  They
will get a potential at (open-string) one-loop from the
interaction with the $N_f-N_c$ D4 branes; this agrees with the
fact that the field theory one-loop potential for $\delta\Phi_0$
is proportional to $N_f-N_c$. This dependence of $\nf -\nc$ was
quite curious from the field theory perspective, but appears
obvious from the point of view of the brane
construction.\foot{More difficult to see in the brane construction
are the other pseudo-moduli, as well as the Goldstone bosons from
the spontaneously-broken flavor symmetry. Since these correspond
to compact field excitations, we expect that they are localized at
the singularity where the two stacks of D4 branes meet the D6
branes.}

\newsec{The M-theory Lift}

In the previous section, we constructed a SUSY-breaking
configuration of D4, NS5 and D6 branes at infinitesimal $g_s$, and
we argued that this is an extrapolation of the SUSY-breaking
vacuum of SQCD to classical string theory. However, it remains to
be seen whether this brane configuration survives the
extrapolation to nonzero $g_s$, i.e.\ whether it remains
(meta)stable once string interactions are taken into account. In
this section, we will analyze the effects of these interactions by
lifting our brane configuration to M-theory.

\subsec{Taub-NUT coordinates}

At nonzero $g_s$, the system is described by M-theory on a circle
of radius
$R=g_s\ell_s$. The D6 branes, which are extended in the 0123789
directions, lift to Taub-NUT (TN) space with asymptotic radius $R$
and charge $N_f$. All the other branes lift to M5 branes extended
in the 0123 directions and wrapping various complex curves inside
$\CM_6= {\rm TN}\times {\Bbb R}^2$, where the second factor
represents the 89 directions
\refs{\WittenEP\HoriAB-\BrandhuberIY}.  An M5 brane
is supersymmetric if and only if it wraps a complex curve that is
holomorphic with respect to a complex structure of $\CM_6$. Several M5
branes form a supersymmetric configuration if and only if they wrap
curves holomorphic with respect to the same complex structure.  We
will find it convenient at various points to parameterize $\CM_6$
using two different coordinate systems. Let us now describe these
coordinate systems in detail.

The first coordinate system we shall refer to as the ``physical
coordinates'' because they have a natural reduction to IIA. In
these coordinates, Taub-NUT is parameterized by $(\vec
r,x^{10})=(x^4,x^5,x^6,x^{10})$; and ${\Bbb R}^2$ by $(x^8,x^9)$
plane. The metric is
\eqn\metric{\eqalign{ ds^2 & =G_{AB}dX^A dX^B
  = V
d{\vec r}^{~2}+V^{-1}(dx^{10}+\vec\omega\cdot d\vec r~)^2 +
(dx^8)^2+(dx^9)^2 }}
where
\eqn\Vdef{ V = 1+{N_f R\over r} }
and
\eqn\omegadef{
 \nabla\times\vec\omega=\nabla V
}
For reasons that will be clear in a moment, we will choose to work
in the gauge where
\eqn\omegaval{ \vec\omega={N_f R\over r(r+x^6)}(-x^5,x^4,0) }

The second coordinate system we will refer to as the ``holomorphic
coordinates" because these make explicit the complex K\"ahler
structure of $\CM_6$. In these coordinates, Taub-NUT is described
by complex variables $(v,\,y)$; and ${\Bbb R}^2$ by $w$. Our
conventions will be that $(v,y,w)$ have mass dimensions given by
\eqn\massdim{
[v]=1,\qquad [y]=2N_c,\qquad [w]=2
}
respectively. As described in \refs{\GibbonsNT,\HitchinZR}
(see also \HoriIW\ and the
appendix of \deBoerZY),
the explicit change of
coordinates between the physical and the holomorphic coordinates
is
\eqn\coordyv{ v={x^4+ix^5\over \ell_s^2};\qquad y=\mu^{2N_c}e^{(x^6-L_0+ix^{10})/2R}\left({r+x^6\over
R}\right)^{N_f/2};\qquad w={x^{8}+ix^9\over R\ell_s^2} }
Here $L_0$ is some normalization constant and $\mu$ is some
arbitrary dimensionful scale required to give $y$ the correct
dimensions. In terms of these, the metric becomes
\eqn\metricyv{ ds^2=G_{i\bar j}dX^i dX^{\bar j}=\left(1+{N_fR\over
r}\right)|\ell_s^2dv|^2+R^2\left(1+{N_fR\over
r}\right)^{-1}\left|f {dv\over v}-2{dy\over
y}\right|^2+R^2|\ell_s^2dw|^2 } where \eqn\Fdef{
f=N_f\left(1-{x^6\over r}\right) }

\subsec{Supersymmetric brane configurations in M-theory}

In this subsection, we will review the supersymmetric M5 brane
configurations constructed by \refs{\HoriAB,\BrandhuberIY}. These
are holomorphic with respect to the complex structure described by
the coordinates $(v,w,y)$. First we consider the curve that
describes the origin of the moduli space of $m=0$ SQCD. It has two
components
 \eqn\uniquesolmuii{\eqalign{
 & \CC_{NS}: \qquad w(z)= 0,\qquad v(z)=z,\qquad
 y(z)=\Lambda^{3N_c-N_f}z^{N_f-N_c}\cr
 & \CC_{NS'}:\qquad w(z) = z,\qquad v(z)=0,\qquad
y(z)=z^{N_c}}}
where we have denoted by $\Lambda$ the dynamical scale of the
theory. The $\CC_{NS}$ component of the curve describes the
$N_f-N_c$ D4 branes ending on the NS brane, while the $\CC_{NS'}$
component of the curve describes the $N_c$ D4 branes ending on the
NS' brane.

When the mass is nonzero, the supersymmetric curve has only one
component and takes the form:
\eqn\mecurve {w=z; \qquad v=m{z+z_0 \over z}; \qquad
y={(z+z_0)^{N_f}\over z^{N_f-N_c}}.}
Here $m$ corresponds to the mass of the electric quarks, while
$z_0^{N_c}=m^{N_f-N_c} \Lambda^{3N_c-N_f}$.

From these formulas, it is straightforward to extract the
behavior of the supersymmetric curves in various limits.

\lfm{1.} $v\to\infty$, leads to \eqn\bcihol{
 w\to 0,\qquad y\to \Lambda^{3N_c-N_f}v^{N_f-N_c}+\dots
 }
We will refer to this  as the NS asymptotic region.
\lfm{2.} $w\to \infty$, leads to \eqn\bciihol{
 v\to m,\qquad y\to w^{N_c}+\dots
}
We will refer to this as the NS' asymptotic region.
\lfm{3.} In addition,
the nature of the map \coordyv\ between the holomorphic and the
physical coordinates requires the condition:
\eqn\conddsix{y=0 ~~~{\rm only ~if}~~~ v=0.}
If this condition were violated, the
curve would have a third, unphysical asymptotic region where the
M5 branes stretch to $x^6\to -\infty$ (see \coordyv).

\medskip
For completeness we note that, as mentioned in the introduction,
using equation \uniquesolmuii\ for the holomorphic massless curve
we can derive two brane configurations: one describing the
electric setup, and the other the magnetic setup
\refs{\SchmaltzSQ,\HoriIW}. This is done by writing the two
components of the curve \uniquesolmuii\ in terms of physical
coordinates \coordyv\ and taking the limit $R \to 0$. The same is
not true in the case when $m\ne 0$: the one component curve
\mecurve\ reduces, upon taking $R\to 0$, to the supersymmetric
electric brane configurations, but fails to give the magnetic one.

\subsec{Non-supersymmetric brane configuration -- no smooth
solution}

Now we are finally ready to tackle the question of whether the
non-supersymmetric brane configuration of the previous section
survives the continuation to non-zero $g_s$. Such a configuration
would be an M5 brane wrapping a non-holomorphic, minimal area
surface in the (Taub-NUT)$\times{\Bbb R}^2$ background, and
satisfying the boundary conditions of the previous subsection.

We will have to simplify things by assuming the following ansatz:

\lfm{A.} The curve has two components corresponding to the NS5 and
NS5' regions, just as in the case of the $m=0$ supersymmetric curve.
These two components touch at the D6 brane (remember, this is at
$x^{4,5,6}=0$). Only the NS5' component is non-holomorphic and
different from the $m=0$ curve. This ansatz is motivated by the
picture of the non-supersymmetric IIA brane configuration (see
figure 7), where only the NS5' is deformed by turning on $m$. It is
also justified by the expectation that the curve should have a
singularity at the D6 brane, since there must be massless modes (the
Goldstone bosons and some of the pseudo-moduli) supported there. By
contrast, the one-component supersymmetric curve is smooth
everywhere, and correspondingly, the supersymmetric vacuum is
gapped.

\lfm{B.} $x^6+ix^{10}$ with $x^6\ge 0$ is a good, global
coordinate along the NS' component. Moreover, we will take $m$ to
be real without loss of generality, and then we will assume that
only $x^4$ is nonzero along the curve. These assumptions are
motivated by the picture of the non-supersymmetric IIA brane
configuration, and they greatly simplify the analysis.\foot{In
fact, one can make even a weaker ansatz that does not assume
$x^6\ge 0$ is a good coordinate on the entire M-theory curve; see
below.}

\lfm{C.} Finally, we assume that the only $x^{10}$ dependence of
the curve comes in through the phase of $x^8+ i x^9$, as
\eqn\argrel{
{\rm arg}(x^8+ix^9)={x^{10}\over 2N_cR }
}
This assumption is justified by the $U(1)$ symmetries of the
problem. The meta-stable state of SQCD has an accidental $U(1)_R$
symmetry that is only broken by the anomaly \iss. In the brane
configuration, this $U(1)_R$ is identified with rotations in the
89 plane, as discussed above in section 2.4, while the anomaly is
identified as a shift in $x^{10}$ under rotations in the 89 plane
\WittenEP, given precisely by \argrel.

\medskip

To summarize, our ansatz restricts the form of the non-holomorphic
curve to be:
\eqn\pertsol{ x^4=f(s),\qquad x^5=0,\qquad x^8+ix^9=e^{ix^{10}/2N_c
R}g(s),\qquad x^6=s }
with $s\ge 0$ and $f$ and $g$ real. Now, the action of such a
surface is simply
\eqn\actionfull{ A = \int d^2z \sqrt{\det g_{ab}} }
where $g_{ab}$ is the $2\times 2$ induced metric
\eqn\indmet{ g_{ab} = {\partial X^A\over\partial z^a} {\partial
X^B\over\partial z^b}G_{AB}(X) } By substituting \pertsol\ and
\metric\ into the action \actionfull\ we find the action:
\eqn\actionrrew{\eqalign{
 &A=\int ds\,
 \sqrt{\left(V^{-1}+{g^2\over 4R^2N_c^2}\right)
 \Big(V(1+f'^2)+g'^2\Big)}\cr
}} where $V$ is the harmonic function of the Taub-NUT space sourced
by the $\nf$ D6 branes: \eqn\rdef{ V = 1+ { N_f R \over
\sqrt{f^2+s^2}} } Using this action we can derive the
Euler-Lagrange equations for $f$ and $g$. The equations of motion
are quite complicated, and we will not show them here. However, we
were able to solve them exactly for $f$; the answer is
\eqn\fsolexact{ f = {c_2+c_3s+c_1\sqrt{(c_2+c_3
s)^2+(1-c_1^2)s^2}\over 1-c_1^2}
 }
where $c_{1,2,3}$ are integration constants. The derivation of
\fsolexact\ is left to appendix A. As a consistency check, note
that a straight line $f(s)=as+b$ is always a solution. Such a line
reduces in type IIA to a D4 in the $x^4-x^6$ plane, which is
always BPS with respect to the D6 branes and the NS' brane. So we
see that the M-theory equations of motion have correctly
reproduced the BPS configurations expected from a IIA analysis.

Now let us apply the boundary conditions $f(s\to \infty)\to \Delta
x$ and $f(s=0)=0$ to determine the integration constants
$c_{1,2,3}$. Here we will run into a problem. The boundary
condition at infinity requires $c_3=-c_1$ and $c_2=\Delta x$, but
then $f(s=0)={\Delta x\over 1-c_1}\ne 0$, so the M5 brane does not
end on the D6 brane. Let us illustrate this point with the
simplest example, namely $c_1=c_3=0$ and $f(s)=\Delta x$. Then the
equation of motion for $g(s)$ (see appendix A) can be solved
exactly; after a straightforward calculation, one finds
\eqn\fconstsol{ g(s) =R\ell_s^2\mu^2 e^{(s-L_0)/2N_c R}\left(
{s+\sqrt{(\Delta x)^2+s^2}\over R}\right)^{N_f/2N_c} }
Note that all of the integration constants in \fconstsol\ have been
fixed by the boundary condition 2 of section 3.2, i.e.\ $y\to w^{N_c}$. In
fact, in holomorphic coordinates, our solution \fconstsol\ is simply
$v=m$, $y=w^{N_c}$ -- it is the simplest solution of boundary
condition 2. Unfortunately, it violates boundary condition 3 of
section 3.2, since it does not end on the D6 brane.

In any event, the general solution \fsolexact\ proves a somewhat
surprising result: there is no smooth, non-holomorphic M-theory
curve obeying the ansatz \pertsol, which satisfies the boundary
conditions at infinity and ends on the D6 brane at $s=0$ !

Let us understand this conclusion in more detail.  For vanishing
$g_s$ the theory is specified by its brane construction.  For
nonzero $g_s$ this specification of the theory is not precise for
two reasons. First, the branes bend at infinity and we should
specify the proper boundary conditions. Second, the finite branes
in the interior become dynamical and should not be specified;
their positions should be set dynamically.  Now, let us examine
the possible boundary conditions for the brane configuration of
figure 7.  There are two reasonable options:
 \item{1.} We can take the same boundary conditions as the asymptotic
 behavior of the M-theory curve \mecurve.  This guarantees that we
 study another state in the same theory.  As we showed, there
 is no solution with these boundary conditions and therefore MQCD
 does not have such a state.
 \item{2.} We can let the M-theory lift of the brane configuration
 of figure 7 bend along the direction of the D4-branes which end
 on it.  Clearly, such an M-theory curve exists.  It has two
 components which preserve different supersymmetries.  Hence, the
 whole curve is not supersymmetric.  The state of M-theory which
 corresponds to this M5-brane configuration might or might not be
 stable.  But in any event, since it has different boundary
 conditions, it is not in the Hilbert space of MQCD.

\medskip

Let us expand on the second option in more detail. Suppose that
instead of first imposing the boundary condition at $s\to \infty$,
we first require the M5 branes to end on the D6 branes
($f(s=0)=0$). This implies that the solution is a straight line
\eqn\linef{
f(s)=c~s
}
and this will never have the correct asymptotics as $s\to \infty$.
In particular, it will not preserve the same supersymmetries as
the NS component of the curve. If $f(s)=c~s$, then the equation of
motion for $g$ is solved by
\eqn\gsolfline{
g(s) = {1\over \tilde\mu}e^{\sqrt{1+c^2}(s-L)/ 2N_c
R}\left({s\over R}\right)^{N_f/2N_c}
 }
where $\tilde \mu$ and $L$ are arbitrary (dimensionful)
parameters. In fact, if we choose $L=L_0$ and $c=\Delta x/L_0$,
then in the $R\to 0$ limit, this straight-line solution will
reduce to the IIA brane configuration depicted in figure 7. So
what we have found is an M-theory lift of the IIA brane
configuration with behavior at infinity {\it different} than the
SUSY curve \mecurve. Since different boundary conditions at
infinity label different Hilbert spaces, our analysis implies that
the SUSY brane configuration and the SUSY-breaking brane
configuration are vacua of {\it different theories}. This should
be contrasted with the field theory, where the SUSY and metastable
states are vacua of the {\it same} theory.

To conclude this section, let us briefly explain how relaxing the
ansatz \pertsol\ to allow for $x^6$ to be of indefinite sign does
not help in obtaining a smooth curve corresponding to the
meta-stable vacuum. A more general ansatz than  \pertsol\ can be
written by taking $w$ as the coordinate along the NS' curve.  The
non-holomorphic NS' component, at general $m$ and $R$, takes the
following form:
\eqn\nonsusyansatz{
 x^4=x^4(|\tilde w|^2),\qquad x^5=0,\qquad x^6=x^6(|\tilde w|^2),\qquad x^{10}=2 R N_c \arg \tilde w
} where $\tilde w \equiv x^8+ix^9$. Now $x^6$ can be both positive
and negative; however, we are assuming implicitly that the complex
$\tilde w$ plane is a good global cover of the entire NS'
component. Substituting this ansatz into the action \actionfull,
with $z=\tilde w$, we find that ${\rm arg}\tilde w$ drops out,
leaving an integral over $\tilde r\equiv |\tilde w|^2$. In
addition, the action has a global $U(1)$ invariance of rotations
in the $(x^4,\,x^6)$ plane. By changing variables to polar
coordinates $(x^4,\,x^6)\to(h(\tilde r),\,\theta(\tilde r))$, one
can show that the action depends only on the derivative of
$\theta(\tilde r)$ and not on $\theta(\tilde r)$ itself. Therefore
the variation with respect to $\theta$ produces an integration
constant, which can be shown to be $\sim \theta'^2$ and vanishes
only if $\theta'=0$. In fact it is not hard to see that our new
ansatz, together with the requirement that the curve ends on the
D6 brane, requires this integration constant to be identically zero.
Therefore in agreement with previous discussion we conclude that
the only solution is a straight line in the $(x^4,\,x^6)$ plane.
This again violates the boundary condition that $x^4\to \Delta x$
as $x^6\to \infty$.

\newsec{IIA Brane Configurations at $g_s\ne 0$}

In the previous section, we saw how there was no M-theory lift of
the meta-stable state of SQCD. Since the calculations done there
involved the classical worldvolume action of the M5 brane, they
were, strictly speaking, only valid in the large $R=g_s\ell_s$
limit. Thus, one might still wonder whether there exists a
meta-stable state at small $R$. In this brief section, we will
present some heuristic arguments indicating that such a state does
not exist even for small $R$.

The analysis is quite similar to the one of the previous section.
Instead of using the M5 brane action, we use the effective action
of a type IIA NS5 brane, and describe the D4's ending on it as a
spike in the NS5 worldvolume.\foot{This is very similar to the way
in which one describes strings ending on a D-brane using the DBI
action \CallanKZ.} We will investigate the SUSY-breaking brane
configuration in a regime where $g_s$ and $l_s$ are small, and
$\Delta L$, $L_0$ and $\Delta x$ are all of order one, much larger
than $l_s$. In this regime the NS5 and D6 are far away and do not
considerably influence the dynamics on the NS5' worldvolume.

Just as at the end the previous section, it will be more
convenient here to parameterize the worldvolume of the NS5 brane
with $(x^8,\,x^9)$ instead of $(x^6,\,x^{10})$. So the relevant
part of the NS5' worldvolume Lagrangian \nsact\ is
\eqn\dbi{
S = \int dx^8 dx^9\, g_s^{-2}\sqrt{\det\left( g_{mn}+g_s^2\,F_m
F_n\right)}
}
where $m$, $n$ range over $89$; and $g_{mn}$ and $F_m$ are the
induced metric and the one-form field strength, respectively, on
the NS5' worldvolume. The field strength is sourced by the $N_c$
D4 branes ending on the NS5' brane; in polar coordinates
$(x^8,x^9)\to (r,\theta)$, we have
\eqn\Ftheta{
F_\theta = N_c \ell_s
}
Since the D4 branes are stretched in the 46 directions, we will assume
that $x^{4,6,8,9}$ are the only scalars excited on the NS5'
worldvolume action. Moreover, we will assume that the solution has
radial symmetry, so $x^M=x^M(r)$. These assumptions stem from the same
physical arguments used to justify the M-theory ansatz in subsection
3.3. The Lagrangian now becomes
\eqn\dbiii{
L = \sqrt{ ( 1+(\dot{x}^6)^2+(\dot{x}^4)^2)(r^2+Q^2)}
}
where the dot denotes differentiation with respect to $r$, and
\eqn\Qdef{
 Q = N_c g_s\ell_s
 }
The equations of motion are easily solved for this system; the
most general solution is
\eqn\xfxssol{\eqalign{
 & x^4(r) = c_3+ c_1\log\left( r+\sqrt{r^2+Q^2-c_1^2-c_2^2}\right)\cr
 & x^6(r) = c_4 + c_2\log\left( r+\sqrt{r^2+Q^2-c_1^2-c_2^2}\right)
}}
with $c_{1,2,3,4}$ integration constants.

As we have discussed above, our candidate for a meta-stable vacuum
should have the same boundary conditions as the brane
configuration corresponding to the massive supersymmetric vacuum,
because both are vacua of the same theory. It is straightforward
to obtain the asymptotic behavior of the supersymmetric vacuum by
reducing the massive M-theory curve; the result is (keep in mind
that we are neglecting the effect of the D6 branes)
\eqn\sol{
x^4\to m,\qquad x^6\to Q \log\left( {r\over r_0}\right)
}
as $r\to \infty$. Here $r_0$ is some length scale, required on
dimensional grounds. Comparing with \xfxssol, we see that the
theory on the NS' brane has the same problem as the M-theory curve
of the previous section -- there is no solution $x^4(r)$, $x^6(r)$
to the equations of motion that satisfies \sol\ and intersects the
D6 brane at $x^4=x^6=0$. Therefore, we conclude that even for
small $g_s$, there is no meta-stable brane configuration
satisfying all the desired boundary conditions.

\newsec{Kink quasi-solutions}

In the previous two sections, we have argued that at $g_s \neq 0$
there is no non-supersymmetric brane configuration corresponding
to the meta-stable vacuum of SQCD. However, the question still
remains: what happens to the IIA brane configuration of figure 7
when $g_s$ is increased? One answer is that it becomes the curve
\linef-\gsolfline\ with the wrong behavior at infinity. On the
other hand, if we fix the boundary conditions at infinity, then
the IIA brane configuration must develop an instability somewhere.

In this section, we will analyze this instability in some detail,
by constructing a kink ``quasi-solution" that satisfies the
equations of motion everywhere except along a co-dimension one
surface, has the right boundary conditions, and reduces to the
meta-stable configuration as $g_s \rightarrow 0$. We will identify
a runaway tadpole mode in this solution, which destabilizes it at
nonzero $g_s$. In the $g_s\to 0$ limit, this tadpole mode is
frozen out -- its kinetic term diverges by a non-perturbative
amount.

One can construct these kink quasi-solutions either using the M5 brane
action in M-theory, or using the type IIA NS5 worldvolume action with
a D4 spike, that we have used in section 4.  In the following two
subsections we analyze these two setups and show that their physics is
identical. In the last subsection we analyze analytically the runaway
mode in the IIA construction.

\subsec{An M-theory kink }

To build a kink solution in M-theory with the right boundary
conditions, the simplest thing to do is to patch together the
straight-line curve \linef-\gsolfline, which has the correct
behavior at the D6 brane, together with the solution \fconstsol,
which has the correct behavior at infinity. The result is a family
of kink quasi-solutions,
\eqn\kinkgen{\eqalign{
 x^6<L:&\qquad |x^8+ix^9| = \cr
 &\qquad\qquad R\ell_s^2\mu^2  e^{-(L_0-L)/2N_c R}e^{
\sqrt{1+(\Delta
x/L)^2}(x^6-L)/2N_cR}\left({\left(1+\sqrt{1+(\Delta
x/L)^2}\right)x^6\over R}\right)^{N_f/2N_c}\cr
 & \qquad x^4 = {\Delta x\over L} x^6\cr
 x^6>L:&
   \qquad |x^8+ix^9| =  R\ell_s^2\mu^2 e^{
 (x^6-L_0)/2N_cR}\left({x^6+ \sqrt{(x^6)^2+(\Delta x)^2}\over
 R}\right)^{N_f/2N_c}\cr
  &\qquad x^4= \Delta x
 }}
parameterized by the position $x^6=L$ of the kink. Substituting
into the action and integrating, we find after a few technically
involved but straightforward steps
\eqn\actdiff{
S(L)-S(L_0) = \Big(\sqrt{L^2+(\Delta
x)^2}-L\Big)-\Big(\sqrt{L_0^2+(\Delta x)^2}-L_0\Big)
}
As expected, this potential has runaway behavior which pushes
$L\to \infty$.

\subsec{A IIA kink }

Since the tadpole found above exists arbitrarily close to $g_s=0$
it is also instructive to present the kink quasi-solutions using
the IIA NS5 worldvolume action. Starting from the general solution
\xfxssol\ and using the same strategy as in the M-theory curve, we
obtain \eqn\solkinkred{\eqalign{
 x^6<L:\qquad & x^4 = \Delta x+{Q\Delta x\over \sqrt{(\Delta x)^2+L^2}}\log{r\over
 r_{kink}}\cr & x^6 = {L\over \Delta x}x^4 \cr x^6>L:\qquad & x^4=\Delta x\cr &
 x^6=L_0 + Q\log {r\over r_0} }
}
where $r_{kink}=r_0\exp\left( { L-L_0\over Q}\right)$ describes
the point in $r$ that the two curves meet and $Q$ is defined in
\Qdef. The curve meets the D6 brane at
$r_{D6}=r_{0}e^{(L-L_0-\sqrt{(\Delta x)^2+L^2})/Q}$. Substituting
the curve into the NS5 Lagrangian \dbiii\ and integrating from
$r_{D6}$ to some cutoff, we find the action of the kink:
\eqn\actionkinkiii{
S = V(L) \equiv  Q\left(\sqrt{L^2+(\Delta x)^2}-L\right)-{1\over
2}\exp\left[{-2 \left( \sqrt{L^2+(\Delta
x)^2}-L+L_0\right)/Q}\right]
}
One can check that this potential has no minimum, and leads to
runaway behavior $L\to\infty$. In the limit $Q\to 0$ the second
term is non-perturbatively small and is negligible relative to the
first term. The M-theory calculation only gave us the first term
in the right hand side of \actionkinkiii; thus the second term
probably comes from the small mistake we are making in neglecting
the effect of the D6 brane.

\subsec{Effective 4D field theory of the kink}

In order to understand the physics of the kink quasi-solution
\solkinkred, it is very instructive to find the effective 4D field
theory that describes its runaway mode.  To do that, we promote
$L$ to a field, $L=L(x^\mu)$ where $\mu=0,1,2,3$. Then the kink
quasi-solution \solkinkred\ depends on $x^\mu$ through $L$. The
effective 4D field theory of the kink can be analyzed both using
the M-theory kink and the IIA one. They both give the same result.
Since the M-theory calculation involves complicated integrals, and
is less clear than the IIA calculation, we only present the
latter.

We substitute the kink into the full NS5' action,
\eqn\fullaction{ S = \int d^4x\,\int dr\,d\theta\,
\sqrt{\det(g_{ab}+g_s^2F_a F_b)} } and expand up to quadratic
order in derivatives of $L(x^\mu)$. The action now involves all of
the NS5' worldvolume coordinates \eqn\wvcoord{
x^a,\,x^b\in(x^\mu,\,r,\,\theta) } while the induced metric is now
$g_{ab}=\partial_a x^M
\partial_b x^M$, with $M$ running over the space-time
coordinates \eqn\stcoord{ x^M\in
(x^\mu,r,\theta,x^4(r;L(x^\mu)),x^6(r;L(x^\mu))) } The result of
the expansion is the following 4D effective action:
\eqn\actionexpandder{ S = \int d^4x\, \left[ V(L(x^\mu)) +
f(L(x^\mu))(\partial_\mu L)^2 + \dots\right] } where $V(L(x^\mu))$
is defined in \actionkinkiii\ and
\eqn\fkindef{\eqalign{
 & f(L)= {1\over6}Q \Delta x \sin\theta -
e^{-2(L_0+\Delta x\tan{\theta\over2})/Q}\left(\sin^4{\theta\over
2}+{Q^2\over 8(\Delta x)^2}\sin^4\theta\right) \cr
 &\qquad \qquad +
e^{2(L-L_0)/Q}\left(
 \sin^2{\theta\over2}-{Q\over
 4\Delta x}\sin^3\theta+{Q^2\over8(\Delta x)^2}\sin^4\theta\right)\cr
  & \sin\theta \equiv {\Delta x\over\sqrt{(\Delta x)^2+L^2}},\qquad 0\le \theta\le
  {\pi\over2}
}}
So the coefficient of the kinetic term of $L$ has very different
$g_s\to 0$ limits depending on whether $L<L_0$ or $L>L_0$. When
$L>L_0$ it diverges due to the exponential factor
$e^{2(L-L_0)/Q}$. When $L<L_0$ it is regular, since both
exponential factors are very small. Therefore, in the $g_s\to 0$
limit, we have a mode which sees the potential \actionkinkiii\ for
$L<L_0$, rolls to $L=L_0$, and then freezes out because it
acquires an infinite kinetic term for $L>L_0$. This reproduces
exactly the $g_s=0$ configuration discussed in section 3.

As $g_s$ starts increasing, the kinetic term for the kink becomes finite
everywhere, and the kink runs away to infinity. Even if the kink
is not an exact solution, we believe this frozen runaway mode
closely approximates the physics of the meta-stable configuration
at $g_s \neq 0$.

\bigskip

\noindent{\bf Acknowledgments}

\noindent We would like to acknowledge discussions with the
participants in the IAS string theory journal club of May 3 2006,
where we found the $g_s =0$ SUSY-breaking brane configurations in
figures 6 and  7. We are indebted to Ken Intriligator for interesting
discussions and early collaboration on this project. We would also
like to thank David Kutasov, Gregory Moore, Raul Rabadan and Edward
Witten, for useful discussions.

The research of IB is supported in part by NSF grant PHY-0503584.  The
research of EG is supported in part by DOE grant DOEFG03-97ER40546.
The research of SH and NS is supported in part by DOE grant
DE-FG02-90ER40542.  S.H. is the D. E. Shaw \& Co., L. P. Member at the
Institute for Advanced Study. The research of DS is supported in part
by a Porter Ogden Jacobus Fellowship and by NSF grant PHY-0243680.

\appendix{A}{Exact Results for Minimal-Area M-theory Curves}

In this appendix, we wish to show how to obtain the exact
solutions \fsolexact\ for the non-holomorphic, minimal-area
M-theory curve. The action is
\eqn\actionrrewapp{\eqalign{
 &A=\int ds\,
 \sqrt{\left(V^{-1}+{g^2\over 4R^2N_c^2}\right)
 \Big(V(1+f'^2)+g'^2\Big)}\cr
}}
where for $N_f$ coincident D6 branes, the potential takes the form
 \eqn\rdefapp{ V = 1+ { N_f R \over
\sqrt{f^2+s^2}}
 }
{}From this action, we obtain equations of motion for $f$ and $g$
which are second-order differential equations. Using these, we can
eliminate $g''$ and obtain relatively simple equation involving
only $g$ and $g'$:
\eqn\eomsimpnewapp{ 4N_c^2R^2 \Big(g'^2 \left({\partial V \over
\partial f}-f'{\partial V \over \partial s} \right)+ 2 V^2 f''\Big)-
g^2\Big((1+f'^2) V^2
 \left({\partial V \over \partial f}
-f'{\partial V \over \partial s} \right) - 2V^3 f''\Big)=0 }
%
%
%
Solving for $g'$ and substituting it back (along with its
derivative) into the equations of motion, we find a surprise: $g$
can be eliminated altogether! The resulting equation is:
\eqn\eomvfapp{f'''\left({\partial V \over \partial f} - f' {\partial V
\over \partial s} \right) + f''\left(3 f'' {\partial V \over
\partial s} + (f'^2-1) {\partial^2 V \over  \partial f \partial s} +
f' {\partial^2 V \over  \partial s^2 } - f' {\partial^2 V \over
\partial f^2 } \right) = 0 }
One can immediately see from this equation that a straight line
($f''=0$) is always a solution regardless of the form of the
function $V$. Indeed, this is expected from the IIA reduction --
such a line reduces in type IIA to a D4 in the $x^4-x^6$ plane,
which is always BPS with respect to the D6 branes. Even more, the
D4 brane is BPS with respect to an {\it arbitrary} distribution of
parallel D6 branes, which when lifted to M-theory give rise to an
arbitrary $V$. Hence equation \eomvfapp\ correctly reproduces the
BPS configurations of D4 and D6 branes expected from a IIA
analysis. However, these straight line configurations do not give
$f\to m$ as $s\to \infty$, so they have the wrong boundary
conditions at infinity.

To see whether there are any other solutions with the correct
boundary conditions, let us focus for simplicity on the case of
$N_f$ coincident D6 branes. Substituting \rdefapp\ in \eomvfapp,
we obtain a nonlinear differential equation for $f$ alone:
\eqn\feomnewapp{3 f''(f'' s^3- f'  s^2+ f^2 f'' s- f f'^2  s+ f  s+
   f^2 f') -\left(s^2+f^2\right) \left(s f'-f\right) f'''=0
}
In fact, this equation can be integrated in $s$, yielding a second
order differential equation:
\eqn\feomsecondapp{ f'' = b_1 \left( {f-s f'\over
\sqrt{s^2+f^2}}\right)^3 }
where $b_1$ is an integration constant. One can check this, for
instance, by substituting it and its derivative into \feomnewapp.
We can also introduce \eqn\ftdefapp{ \tilde f \equiv {f\over s} }
in terms of which \feomsecondapp\ becomes \eqn\fteomsecondapp{
\tilde f'' + {2\tilde f'\over s}+{b_1 \,s^2 \tilde f'^3\over
(1+\tilde f^2)^{3/2}}=0~.
 }
Amazingly, this equation can also be integrated in $s$, yielding
\eqn\fteomfirstapp{
 \tilde f'-{\sqrt{1+\tilde f^2}\over s^2\left( b_1
\tilde f+b_2\sqrt{1+\tilde f^2}\right)
}=0~. } This equation can be solved exactly, yielding
\eqn\fsolexactapp{ f = {b_2(1-b_3 s)+b_1\sqrt{(1-b_3
s)^2-(b_1^2-b_2^2)s^2}\over b_1^2-b_2^2}
 }
A straightforward change of variables from $b_{1,2,3}$ to
$c_{1,2,3}$ yields the formula \fsolexact\ quoted in the text.

\listrefs

\end